\documentclass[11pt]{article}

\pdfoutput=1

\usepackage{graphicx}

\usepackage{amsmath}
\usepackage{amsfonts}

\begin{document}

\title{Maxwell-type models for the effective thermal conductivity of a porous material with radiative transfer in the voids}

\author{Kristian B. Kiradjiev,$^{a}$ Svenn Anton Halvorsen,$^b$ Robert A. Van Gorder,$^{a*}$\\ and Sam D. Howison$^a$\\    
\small $^a$ Mathematical Institute, University of Oxford \\
\small Andrew Wiles Building, Radcliffe Observatory Quarter, Woodstock Road\\
\small Oxford OX2 6GG United Kingdom\\
\small $^b$ Teknova AS, Tordenskjolds gate 9, 5th floor\\
\small NO-4612 Kristiansand S, Norway\\
\small $^* $Email: Robert.VanGorder@maths.ox.ac.uk}  
\date{\today}       
\maketitle

\begin{abstract}
There are several models for the effective thermal conductivity of two-phase composite materials in terms of the conductivity of the solid and the disperse material. In this paper, we generalise three models of Maxwell type (namely, the classical Maxwell model and two generalisations of it obtained from effective medium theory and differential effective medium theory) so that the resulting effective thermal conductivity accounts for radiative heat transfer within gas voids. In the high-temperature regime, radiative transfer within voids strongly influences the thermal conductivity of the bulk material. Indeed, the utility of these models over classical Maxwell-type models is seen in the high-temperature regime, where they predict that the effective thermal conductivity of the composite material levels off to a constant value (as a function of temperature) at very high temperatures, provided that the material is not too porous, in agreement with experiments. This behaviour is in contrast to models which neglect radiative transfer within the pores, or lumped parameter models, as such models do not resolve the radiative transfer independently from other physical phenomena. Our results may be of particular use for industrial and scientific applications involving heat transfer within porous composite materials taking place in the high-temperature regime.
\end{abstract}


\section{Introduction}
\label{sec:1}

Estimating thermal properties of bulk or composite materials, such as their effective thermal conductivity, is crucial for understanding of heat transfer in them. This on its own is of great significance for various industries which process raw materials, often from a granular or particulate state. There are various models, such as lumped parameter models, which condense the effects of different phenomena such as solid conduction, radiation, and conduction through contact points, into equivalent parameters. See \cite{Kunii1960} and \cite{Kandula2011}, for example, for several expressions for effective thermal conductivity. While useful, such lumped parameter models may obscure the contribution from each phenomenon, and often lack a rigorous foundation. For this reason, it is desirable to derive such models from first principles.

Common models for calculating the effective thermal conductivity of a solid composite material include the Maxwell model (based on the pioneering work of Maxwell \cite{Maxwell1873}, where an expression for the effective thermal conductivity was derived via far-field perturbations to solutions of the steady heat equation), as well as variants thereof, including the effective medium theory (EMT) model and the differential effective medium theory (DEMT) model. EMT uses a similar approach to that used in deriving the Maxwell model, and can be applied to many other physical properties (see \cite{Landauer1952}, for instance, for the electrical resistance problem). In \cite{Carson2005,Xu2016}, there is a comparison between the Maxwell model and EMT, and those works outline how certain bounds on the thermal conductivities can be obtained. The multipole expansion method \cite{Mogilevskaya2012} gives similar results. In applying the DEMT, one incrementally adds one of the materials to the composite, and considers the effect of an infinitesimal change in the composite material composition on the effective thermal conductivity, obtaining a differential equation for the effective thermal conductivity in terms of the volume fraction of inclusions; see \cite{Bauer1993,Ordonez-Miranda2010}. Reviews of many current models and methodologies, including those outlined above, can be found in \cite{Carson2002,Karayacoubian2006,Pietrak2015}. 

Recent work has involved the application of Maxwell-like models to the study of composite materials \cite{xu2016reconstruction}, including polymer composites \cite{kim2017two,zhai2018effective}. Such models have recently proven useful in understanding nanoflake thermal annealing \cite{bernal2017thermally}, and in the understanding of effective thermal conductivity in a variety of materials, such as for a wood cell modelled as a constituent element of briquette chips \cite{sova2018effective}, polyethylene nanocomposites \cite{zabihi2017effective}, phase-change materials  \cite{abujas2016performance} and composites \cite{wang2017experimental}, fiber-reinforced concrete \cite{liu2017theoretical}, alumina-graphene hybrid filled epoxy composites \cite{akhtar2017alumina}, metal-graphene composites \cite{wejrzanowski2016thermal}, transparent and flexible polymers containing fillers \cite{ngo2016thermal}, and composite materials for LED heat sink applications \cite{terentyeva2017analyzing}. Maxwell-like models have also motivated theoretical methods for upscaling the thermal conduction equation in periodic composite materials \cite{mathieu2017method}. The thermal conductivity of composites made up of metallic and non-metallic microparticles or nanoparticles embedded in a solid matrix has been considered in \cite{ordonez2014thermal}, following on from a model for the effective thermal conductivity of metal-nonmetal particulate composites, obtained in \cite{ordonez2012model}. The role of pore shape on the thermal conductivity of porous media has been studied using the Bruggeman differential effective medium theory \cite{ordonez2012effect}.

All of these models were originally derived for composite materials with heat conduction as the only mode of heat transfer. However, radiation can play a strong role in heat transfer in porous media when there is a non-negligible volume fraction of gas voids to solid material. For a review of radiative heat transfer theory, see \cite{Modest1993}. Examples of applications where radiation within the gas voids can influence the effective thermal conductivity of composite materials include multilayer thermal insulation systems \cite{spinnler2004studies}; porous partially stabilised zirconia \cite{hsu1992measurements}; monolithic organic aerogels \cite{lu1992thermal}; ultralight metal foams \cite{zhao2004thermal} and open-celled metal alloy foams \cite{zhao2004temperature}; and Earth materials within the mantle \cite{schatz1972thermal}. Note that many such applications involve high-temperature regimes. The effect of identical approximately spherical pores and anisometric cylindrical pores on the thermal conductivity of alumina, graphite, and nickel has been investigated separately in \cite{francl1954thermal}, and pore orientation was shown to affect the value of the effective thermal conductivity for a given porosity, particularly above 500 C. Experimental results for the thermal conductivity of a range of porous materials were observed in \cite{luikov1968thermal}, and for large temperatures thermal conductivities were shown to level off in some cases, rather than to increase without bound. In general, due to the high-temperature regime required in many industries, it is very difficult to conduct experiments, which supports the need for good models for predicting the effective thermal conductivity of composite materials. Some experimental methods for determining this can be seen in \cite{Kosowska-Golachowska2014}, while a review of experimental methods for characterising thermal contact resistance is given in \cite{xian2017experimental}. Loeb \cite{loeb1954thermal} obtained a variety of formulae for thermal conductivity of porous media, which involve the conductivity of the solid material, the emissivity of the surface of the pores, and the size, shape, and distribution of the pores. Loeb \cite{loeb1954thermal} was able to show that materials can be prepared having different thermal conductivities in different directions. A number of correlations between effective thermal conductivity and the packing structure in beds of spherical particles exist in the literature; see, for instance, \cite{van2010review}, for a review on this topic. A homogenization approach, making use of the separation of length scales, was used in \cite{Allaire2014}, where the authors derive an effective thermal conductivity tensor and study the second-order corrections to the temperature field.

In this paper, we generalise effective thermal conductivity models of Maxwell type to include radiation in the gas pores between densely-packed solid particles, justifying the use of an effective radiative conductivity, often as employed in lumped parameter models to account for radiation. In particular, we consider the Maxwell, EMT, and DEMT models, and extend them to the case when we have a solid matrix of material with multiple gas voids, a reasonable assumption for modelling densely-packed particle beds. We allow for thermal conduction in the solid phase and radiation in the gas phase, neglecting thermal conduction in the gas, because its thermal conductivity is relatively small. Similarly, we ignore any convective heat flux in the gas, on the basis that the heat capacity of the gas is also relatively small. 

The remainder of the paper is organised as follows. In Section 2, we review the Maxwell, EMT, and DEMT models for the effective thermal conductivity of a solid matrix of material with multiple inclusions of a second material. In Section 3, we extend the Maxwell model to include radiation in the gas phase, deriving a new effective thermal conductivity. In Sections 4 and 5, respectively, we derive EMT and DEMT models for effective thermal conductivity, accounting for radiative transfer in the gas phase. In Section 6, we compare and discuss the results of these three models. We conclude in Section 7.

\section{Effective thermal conductivity models of Maxwell type}
\label{sec:2}

\begin{figure*}
\begin{center}
\includegraphics[width=0.3\linewidth]{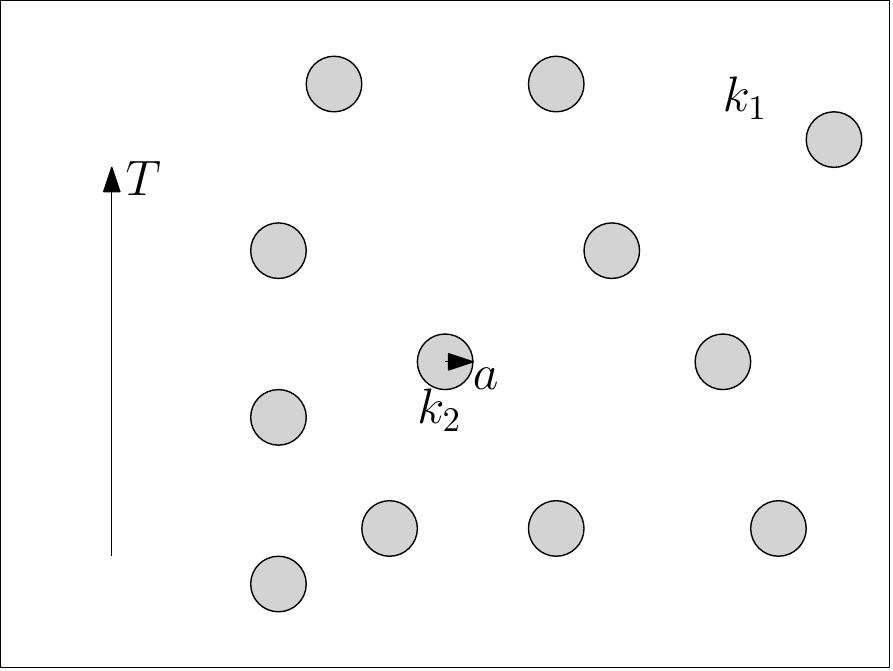}
\includegraphics[width=0.3\linewidth]{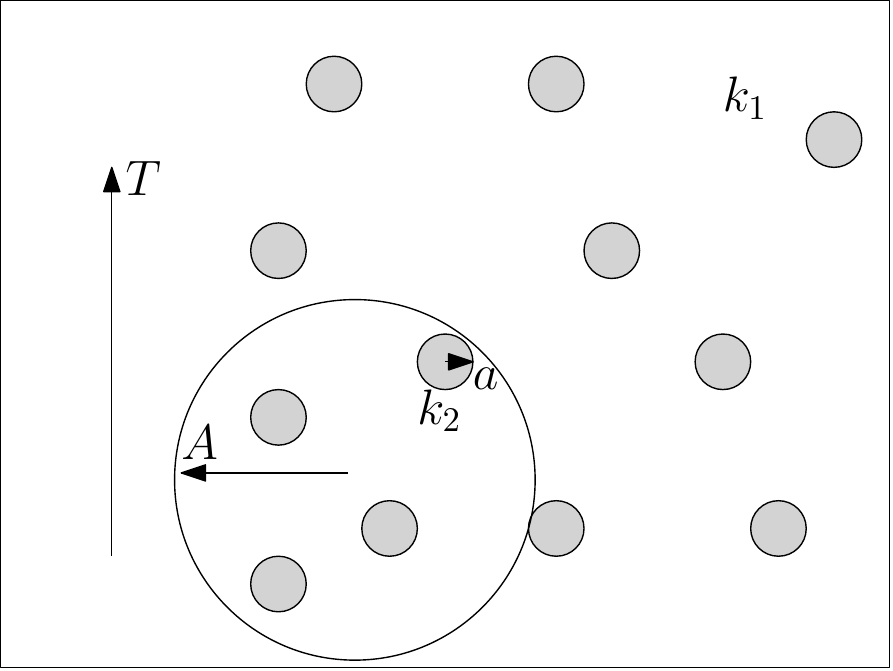}
\includegraphics[width=0.3\linewidth]{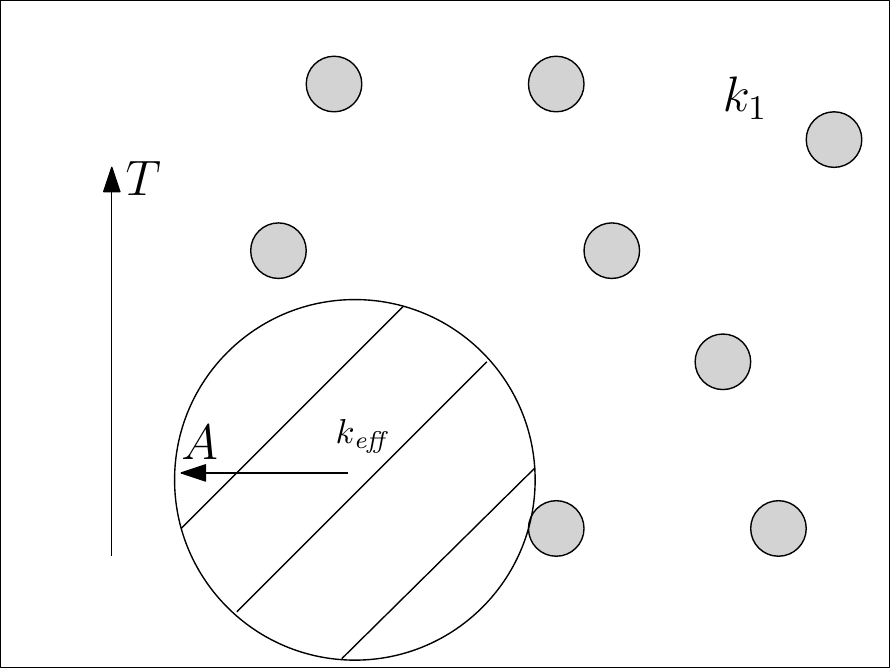}
\caption{Geometric configurations used in Maxwell's model. (a) Initial set-up. (b) A sphere of radius $A$ with spherical inclusions of radius $a$ inside. (c) A sphere of radius $A$ with spherical inclusions of radius $a$, considered as composite material with effective conductivity $k_{\mathit{eff}}$.}
\label{fig:7}
\end{center}
\end{figure*}

Maxwell \cite{Maxwell1873} models the composite material as a continuous matrix of constant conductivity $k_{1}$ containing multiple spherical inclusions of radius $a$ (not necessarily in a regular array) of constant conductivity $k_{2}$ and applies an external temperature field with gradient of magnitude $T'_{\infty}$, which drives heat through the medium (see Figure \ref{fig:7}). Further, he assumes that the sizes of the particle inclusions are small relative to the inter-particle distances, so that the thermal disturbances to the temperature field due to each inclusion can be considered independently.
Consider a single spherical inclusion. To obtain the perturbed temperature field $T_{1}$ and $T_{2}$ outside and inside the sphere, respectively, we solve Laplace's equation with continuity of temperature and conductive flux across the interface (ignoring interfacial resistance), and a matching condition with the far-field temperature. Taking spherical coordinates ($r,\varphi,\theta$) with origin at the centre of the sphere, and $\theta=0$ parallel to the imposed temperature gradient, we have
\begin{align}
\nabla^2 T_{1}&=0 \qquad &&\text{ for } \qquad r> a, \label{eq:81}\\
\nabla^2 T_{2}&=0 \qquad &&\text{ for } \qquad r< a, \label{eq:80}\\
T_{1}&=T_{2} \qquad &&\text{ on } \qquad r=a, \label{eq:83}\\
k_{1}\dfrac{\partial T_{1}}{\partial r}&=k_{2}\dfrac{\partial T_{2}}{\partial r} \qquad &&\text{ on } \qquad r=a, \label{eq:91}\\
T_{1} &\to T'_{\infty}r\cos{\theta} \qquad &&\text{ as } \qquad r\to \infty. \label{eq:70}
\end{align}
This is solved to obtain
\begin{align}
T_{1}=\left(T'_{\infty}r+\frac{C_{1}(a,k_{1},k_{2})}{r^2}\right) \cos{\theta}, \label{eq:308}\\
T_{2}=C_{2}(a,k_{1},k_{2})r\cos{\theta},
\label{eq:310}
\end{align}
where
\begin{align}
C_{1}(a,k_{1},k_{2})=\frac{k_{1}-k_{2}}{2k_{1}+k_{2}}a^3T'_{\infty}, \label{eq:309}\\
C_{2}(a,k_{1},k_{2})=\frac{3k_{1}}{2k_{1}+k_{2}}T'_{\infty}.
\label{eq:311}
\end{align}
Here, $C_{1}(a,k_{1},k_{2})$ determines the far-field behaviour induced by the spherical inclusion $a$. Maxwell next proposed to consider a large sphere of radius $A$ in the matrix material with $n$ spherical inclusions of the second material inside it (see Figure \ref{fig:7}). By the assumption that the spherical inclusions are far from each other so that they do not interact, applying the superposition principle, we get that the coefficient, $C_{n}(a,k_{1},k_{2})$, for the far-field behaviour, induced by the $n$ spheres, is simply
\begin{equation}
C_{n}(a,k_{1},k_{2})=nC_{1}(a,k_{1},k_{2})=\frac{k_{1}-k_{2}}{2k_{1}+k_{2}}\phi A^3 T'_{\infty},
\label{eq:241}
\end{equation}
where in the last expression $\phi=n a^3/A^3$ is the volume fraction of the spherical inclusions.

The crux of the method lies in considering the large sphere, which has numerous spherical inclusions inside, as a continuous medium with effective thermal conductivity $k_{\mathit{eff}}$ (see Figure \ref{fig:7}). Using \eqref{eq:309}, the far-field perturbation coefficient $C_{1}(A,k_{1},k_{\mathit{eff}})$ due to a single spherical inclusion with a thermal conductivity $k_{\mathit{eff}}$ is
\begin{equation}
C_{1}(A,k_{1},k_{\mathit{eff}})=\frac{k_{1}-k_{\mathit{eff}}}{2k_{1}+k_{\mathit{eff}}}A^3 T'_{\infty}.
\label{eq:251}
\end{equation}
Since the far-field perturbation is the same both ways, we have
\begin{equation}
C_{n}(a,k_{1},k_{2})=C_{1}(A,k_{1},k_{\mathit{eff}}), 
\label{eq:261}
\end{equation}
and, therefore, rearranging for $k_{\mathit{eff}}$, we obtain
\begin{equation}
k_{\mathit{eff}}=\frac{2k_{1}+k_{2}+2\phi(k_{2}-k_{1})}{2k_{1}+k_{2}-\phi(k_{2}-k_{1})}k_{1}=k_{1}+\frac{3 k_{1} (k_{2}-k_{1})}{2k_{1}+k_{2}}\phi + O(\phi^2).
\label{eq:3}
\end{equation}
This approach gives a good estimate of the effective thermal conductivity in the dilute porosity limit \cite{Pietrak2015}. 
Note that Maxwell's result is consistent with the lower and upper bounds for the effective thermal conductivity of an isotropic medium, derived by other means in \cite{Hashin1962}.

Another approach is that of Effective Medium Theory (EMT), which is used in many other situations involving effective properties such as conductivities, polarisation, and the like. It is derived in a similar way to Maxwell's model, but, crucially, it does not assume an ambient medium solely composed of the matrix material extending to the far-field, where we look for temperature perturbations. However, the assumption for a dilute porosity limit still holds. As one can check in the end, the two formulae agree up to $O(\phi)$ and differ at $O(\phi^2)$ for small $\phi$. 

\begin{figure*}
\begin{center}
\includegraphics[width=0.3\linewidth]{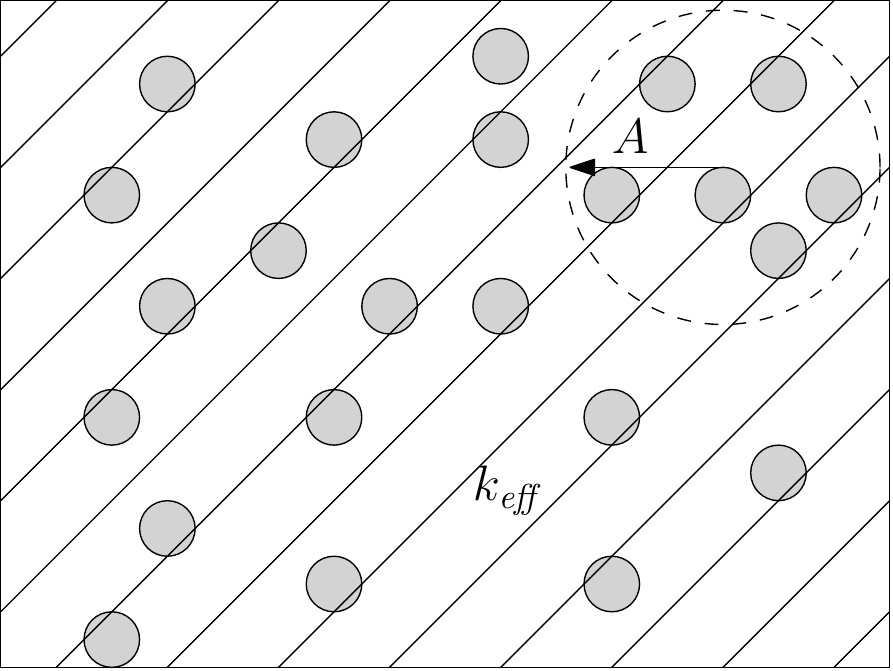}
\includegraphics[width=0.3\linewidth]{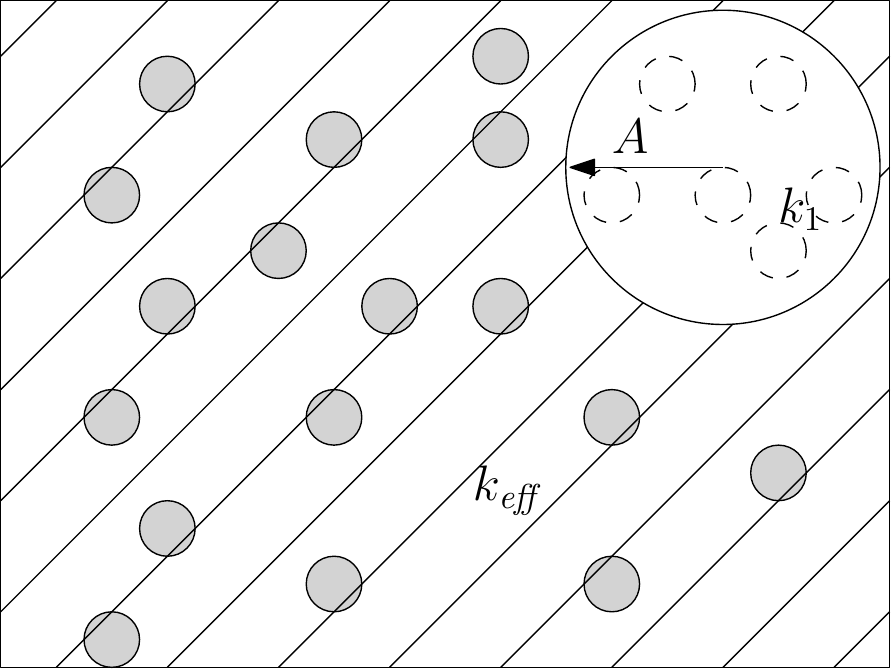}
\includegraphics[width=0.3\linewidth]{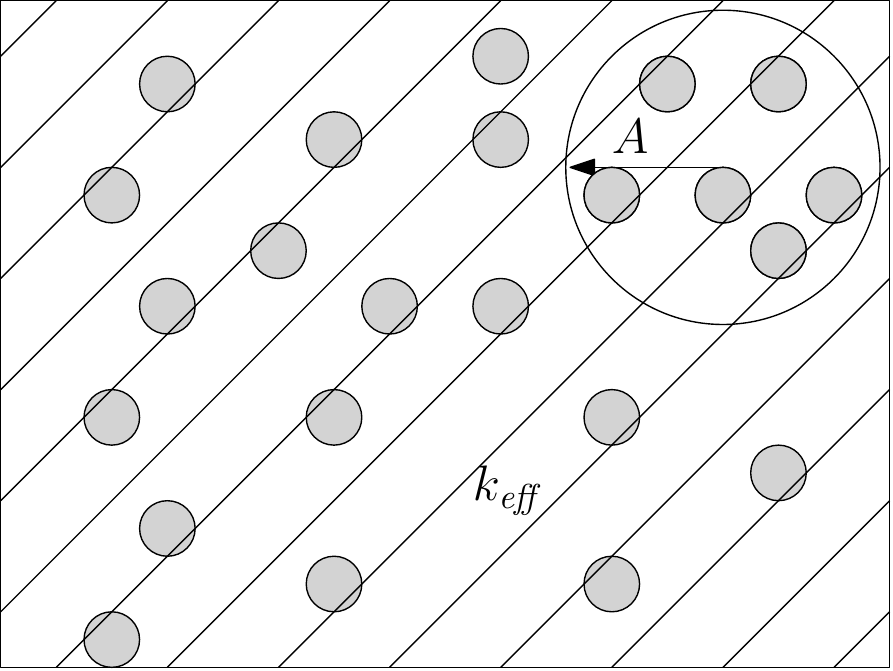}
\caption{Configurations for the EMT model. (a) A random sphere of radius $A$ is selected. (b) The material inside is replaced with that of the matrix. (c) The original spherical inclusions are returned to their place.}
\label{fig:11}
\end{center}
\end{figure*}

In reviewing the effective medium theory model, we follow the approach outlined in \cite{Xu2016}. Unlike Maxwell's model, we begin by treating the composite medium, consisting of spherical inclusions of a different material sufficiently far apart, as a single material with effective conductivity $k_{\mathit{eff}}$. The temperature field of this medium is determined solely by the prescribed temperature gradient and has magnitude
\begin{equation}
T_{1}=T'_{\infty} r\cos \theta.
\label{eq:301}
\end{equation}
Now, as in Maxwell's model, suppose we pick a spherical region of radius $A$ at random (Figure \ref{fig:11}), remove the spherical inclusions (of radius $a$) from it (suppose they are $n$ in number), and replace them with particles of the same conductivity as the matrix material, resulting in a sphere of conductivity $k_{1}$ (Figure \ref{fig:11}). The resulting far-field temperature obtained for a single sphere of radius $A$ immersed in a matrix of thermal conductivity $k_{\mathit{eff}}$ for $r \gg A$ reads
\begin{equation}
T_{1}=\left(T'_{\infty}r+\frac{C_{1}(A,k_{\mathit{eff}},k_{1})}{r^2}\right) \cos{\theta},
\label{eq:312}
\end{equation}
with $C_{1}(A,k_{\mathit{eff}},k_{1})$ defined as in \eqref{eq:251}, noting that the order of the arguments is different this time, because, for this particular calculation, the matrix material has conductivity $k_{\mathit{eff}}$ and the sphere has conductivity $k_{1}$.

We now want to return the original spherical inclusions within the sphere of radius $A$. We first need to vacate $n$ spherical holes of conductivity $k_{1}$, and then replace them with spheres of the second material (Figure \ref{fig:11}). Again, we assume sparsity of the spherical inclusions so that the interaction between them is negligible. The respective contributions to the far-field temperature coefficient are given by
\begin{align}
C_{n}(a,k_{\mathit{eff}},k_{1})=nC_{1}(a,k_{\mathit{eff}},k_{1})&=\frac{k_{\mathit{eff}}-k_{1}}{2k_{\mathit{eff}}+k_{1}}\phi A^3T'_{\infty}, \label{eq:321}\\
C_{n}(a,k_{\mathit{eff}},k_{2})=nC_{1}(a,k_{\mathit{eff}},k_{2})&=\frac{k_{\mathit{eff}}-k_{2}}{2k_{\mathit{eff}}+k_{2}}\phi A^3T'_{\infty}. \label{eq:331}
\end{align}
Combining \eqref{eq:312}-\eqref{eq:331}, we obtain the net far-field temperature
\begin{equation}\begin{aligned}
T_{1} =  \left\lbrace  \vphantom{\frac{C_{1}(A,k_{\mathit{eff}},k_{1})-C_{n}(a,k_{\mathit{eff}},k_{1})+C_{n}(a,k_{\mathit{eff}},k_{2})}{r^2}}T'_{\infty}r  +\frac{C_{1}(A,k_{\mathit{eff}},k_{1})-C_{n}(a,k_{\mathit{eff}},k_{1})+C_{n}(a,k_{\mathit{eff}},k_{2})}{r^2} \right\rbrace\cos{\theta} .
\label{eq:341}
\end{aligned}\end{equation} 
Since we effectively arrive at the initial configuration of homogenised medium with the prescribed temperature gradient, then
\begin{equation}
C_{1}(A,k_{\mathit{eff}},k_{1})-C_{n}(a,k_{\mathit{eff}},k_{1})+C_{n}(a,k_{\mathit{eff}},k_{2})=0,
\label{eq:351}
\end{equation}
which gives an implicit expression for $k_{\mathit{eff}}$:
\begin{equation}
\frac{k_{\mathit{eff}}-k_{1}}{2k_{\mathit{eff}}+k_{1}}(1-\phi)+\frac{k_{\mathit{eff}}-k_{2}}{2k_{\mathit{eff}}+k_{2}}\phi=0.
\label{eq:361}
\end{equation}
This expression is symmetric in $k_{1}$ and $k_{2}$, provided $\phi \leftrightarrow 1-\phi$, as the model can be applied when either of the materials is dilute in the other one. 
Equation \eqref{eq:361} has a unique positive root
\begin{equation}\begin{aligned}
k_{\mathit{eff}} & = \frac{1}{4}\left\lbrace \vphantom{\sqrt{\phi(k_{2})^2}} 3\phi (k_{2}-k_{1})+(2k_{1}-k_{2}) \right. \\
& \qquad \left. +\sqrt{(3\phi (k_{2}-k_{1})+(2k_{1}-k_{2}))^2+8k_{2}k_{1}}\right\rbrace \\ & = k_{1}+\frac{3 k_{1} (k_{2}-k_{1})}{2k_{1}+k_{2}}\phi + O(\phi^2),
\label{eq:5}
\end{aligned}
\end{equation}
for $\phi \ll 1$, which agrees with Maxwell's result up to $O(\phi)$.

A third model, the Differential Effective Medium Theory (DEMT), motivated by the early work of Bruggeman \cite{bruggeman1935dielectric}, considers media of various particle volume fractions \cite{Bauer1993,Ordonez-Miranda2010} but with a range of particle sizes present in the composite medium. In this approach, one incrementally adds one of the materials to the composite, and considers the effect on the effective thermal conductivity, obtaining a differential equation for it in terms of the porosity. 

In reviewing the differential effective medium theory approach, we follow \cite{Ordonez-Miranda2010}. We assume that the effective thermal conductivity is given as a function of the particle volume fraction, $\phi$, resulting in the ansatz
\begin{equation}
k_{\mathit{eff}}(\phi)=k_{1}(1+b(k_{1},k_{2})\phi+c(k_{1},k_{2})\phi^2+\cdots).
\label{eq:410}
\end{equation}
Here $b(k_{1},k_{2})$ determines the behaviour of $k_{\mathit{eff}}$ in the dilute limit $\phi \ll 1$, while $c(k_{1},k_{2})$ is the second-order correction which partially accounts for the particle interactions. 

We remove composite material of volume $\Delta V$, and replace it with the same volume of particle inclusions. Treating the homogeneous material as a new matrix, we may express the new effective conductivity as
\begin{equation}
k_{\mathit{eff}}(\phi+\mathrm{d}\phi) =k_{\mathit{eff}}(\phi)\left(1+b(k_{\mathit{eff}}(\phi),k_{2})\frac{\mathrm{d}V}{V} +\cdots\right),
\label{eq:411}
\end{equation}
where $\mathrm{d}\phi=(\mathrm{d}V-\mathrm{d}V_{p})/V$ is the net increase in the particle volume fraction and $\mathrm{d}V_{p}$ is the volume of the particles removed. Assuming that $\mathrm{d}V_{p}/V_{p}=\mathrm{d}V/V$ on average, we have $\mathrm{d}V/V=\mathrm{d}\phi/(1-\phi)$. Hence,
in the limit $\mathrm{d}\phi\to 0$, we obtain
\begin{equation}
\dfrac{\mathrm{d}k_{\mathit{eff}}}{\mathrm{d}\phi}=\frac{k_{\mathit{eff}}}{1-\phi}b(k_{\mathit{eff}},k_{2}),
\label{eq:423}
\end{equation}
and integrating this differential equation with the condition $k_{\mathit{eff}}(0)=k_{1}$, we find
\begin{equation}
\int_{k_{1}}^{k_{\mathit{eff}}}\frac{\mathrm{d}s}{sb(s,k_{2})}=-\log(1-\phi).
\label{eq:444}
\end{equation}
From Maxwell's result \eqref{eq:3}, we find
\begin{equation}
b(k_{1},k_{2})=\frac{3(k_{2}-k_{1})}{2k_{1}+k_{2}}.
\label{eq:451}
\end{equation}
After performing the integration in \eqref{eq:444}, we obtain the effective thermal conductivity implicitly as
\begin{equation}
\left(\frac{k_{\mathit{eff}}-k_{2}}{k_{1}-k_{2}}\right)^3\frac{k_{1}}{k_{\mathit{eff}}}=(1-\phi)^3.
\label{eq:461}
\end{equation}
This result was obtained (and the cubic equation solved explicitly) in \cite{kamiuto1990examination} (see equations (2) and (3) of \cite{kamiuto1990examination}) and \cite{kamiuto1993combined} (see equation (10) of \cite{kamiuto1993combined}). In order for the manipulations in the derivation of this model to be valid (in particular, to be able to consider incremental changes of the particle volume fractions), particles of a range of sizes are assumed to be present in the composite material, which is usually the case in many industrial processes. We note that if we expand $k_{\mathit{eff}}$ in powers of $\phi$ for small $\phi$, we again obtain agreement with Maxwell's and EMT models up to $O(\phi)$.

As we saw, all three models agree up to $O(\phi)$ in the limit $\phi \to 0$. Maxwell's model is a classical result, which has a surprising accuracy beyond $O(\phi)$ when compared to standard multiple-scales approximations for ordered media \cite{Bruna2015}. The effective medium theory model has an intrinsic symmetry in its constituent materials, which can be used to give estimates of the effective conductivity of composite materials, in which the inclusions are in a continuous rather than discrete phase. The differential effective medium theory model is applicable when there is a gradient in particle size in the composite material, so can be used for heterogeneous materials,  whereas normally Maxwell's and EMT models assume particle inclusions of a uniform size distribution. 

We now extend these three models to the case of a solid matrix of material with multiple gas voids. The assumption of a solid matrix of material with multiple gas voids, a reasonable assumption for modelling densely packed particle beds, which commonly arise in applications in this area. As noted above, we consider conductive heat transfer only via the solid phase. \footnote{The extension to include gas conductivity is straightforward but lengthy.} We do, however, include radiation in the gas phase.

\section{Maxwell model with radiation}
\label{sec:3}

First, we extend the Maxwell model. We consider the case of spherical pores, filled with gas of negligible thermal conductivity. 

\begin{figure}
\centering
\includegraphics[width=0.25\linewidth]{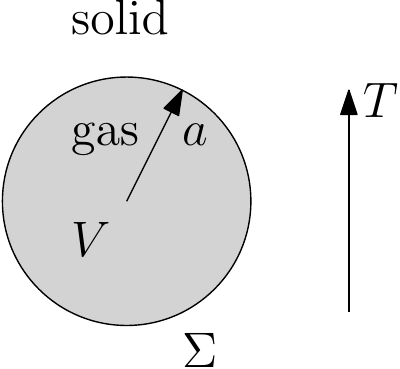}
\caption{Geometry of a spherical void in the Maxwell model with radiation.}
\label{fig:4}
\end{figure}

Similarly to Section \ref{sec:2}, we begin by looking at perturbations to the far-field temperature induced by a single small spherical gas void $V$ of radius $a$ (see Figure \ref{fig:4}) when we apply an external temperature field of constant magnitude $T'_{\infty}$. The energy flux within the void is
assumed to be entirely radiative. Noting that the temperature must be
measured on the absolute scale, by the Stefan--Boltzmann law, the flux
emitted per unit area at a point $\mathbf{r}$ on the void surface $\Sigma$
is $\epsilon\sigma T^4(\mathbf{r})$, where $\epsilon$ is the emissivity of the surface (assumed to be a gray body) and $\sigma$ is the Stefan--Boltzmann constant. The incident flux from points elsewhere on the surface is given by~\cite{Modest1993}
\begin{equation}
\begin{split}
\int_\Sigma  \epsilon\sigma T^4(\mathbf{r}')
\frac{ \cos \gamma(\mathbf{r},\mathbf{r}')
\cos \gamma'(\mathbf{r},\mathbf{r}') \,
\mathrm{d}S'}{\pi|\mathbf{r}-\mathbf{r}'|^2}\\
=\int_\Sigma   \epsilon\sigma T^4(\mathbf{r}') \,
\mathrm{d}F(\mathbf{r},\mathbf{r}'), \label{eq:53}
\end{split}
\end{equation}
where $\gamma$ is the angle between the normal at
$\mathbf{r}$ and the line of sight
from $\mathbf{r}$ to $\mathbf{r}'$,
$\gamma'$ being defined similarly. The term
\begin{equation}
\mathrm{d}F(\mathbf{r},\mathbf{r}')
=\frac{ \cos \gamma(\mathbf{r},\mathbf{r}')
\cos \gamma'(\mathbf{r},\mathbf{r}') \,
\mathrm{d}S'}{\pi|\mathbf{r}-\mathbf{r}'|^2} \label{eq:55}
\end{equation}
is known as the view factor. It is a geometrical property of the domain
and satisfies $\int_\Sigma \mathrm{d}F(\mathbf{r},\mathbf{r}')  = 1$ for
all $\mathbf{r}$. For a sphere of radius $a$, simple trigonometry shows
that the view factor is $\mathrm{d}S'/(4\pi a^2)$.

Denoting the solid-matrix-material conductivity simply by $k$, we now need to solve the following problem:
\begin{align}
\nabla^2 T&=0 \qquad &&\text{ for } \qquad r > a, \label{eq:8}\\
k\dfrac{\partial T}{\partial r}&=\epsilon \sigma T^4-\int_{\Sigma} \frac{\epsilon \sigma T^4}{4\pi a^2} \, \mathrm{d}S \qquad &&\text{ on } \qquad r=a, \label{eq:9}\\
T &\to T'_{\infty} r \cos{\theta} \qquad &&\text{ as } \qquad r\to \infty. \label{eq:10}
\end{align}

As the voids are assumed to be small, the temperature variation across them is also small. Hence, we expand $T$ around some reference temperature $\hat{T}_{0}$, which is taken to be the temperature on the `equator' of the void $\theta=\pi/2$, exploiting the fact that
\begin{equation}
\delta=\frac{aT'_{\infty}}{\hat{T}_{0}}\ll 1.
\label{eq:11}
\end{equation}
We expand $T$ for $r\geq a$ as
\begin{equation}
T \sim \hat{T}_{0}+\delta \hat{T}_{1} + O(\delta^2).
\label{eq:12}
\end{equation}
The $O(1)$ terms cancel and so
\begin{equation}
\delta k \dfrac{\partial \hat{T}_{1}}{\partial r}=\epsilon \sigma \hat{T}_{0}^4+4 \delta \epsilon \sigma \hat{T}_{0}^3\hat{T}_{1}-\int_{\Sigma} \epsilon \sigma (\hat{T}_{0}^4+4 \delta \hat{T}_{0}^3\hat{T}_{1})\mathrm{d}F \text{ on } r=a. \label{eq:13}
\end{equation}
The first term on the left-hand side cancels with the first term in the integral since, by definition,
\begin{equation}
\int_{\Sigma} \mathrm{d}F = 1.
\label{eq:14}
\end{equation}
This simplifies \eqref{eq:13} to
\begin{equation}
\frac{k}{4 \epsilon \sigma \hat{T}_{0}^3}\dfrac{\partial \hat{T}_{1}}{\partial r}=\hat{T}_{1}-\int_{\Sigma} \hat{T}_{1}\mathrm{d}F \qquad \text{ on } \qquad r=a.
\label{eq:15}
\end{equation}
We further note that the integral term in this equation vanishes identically:
all points on the void surface have, to $O(\delta)$, the same incident
flux (but different radiative fluxes). We readily find that
\begin{equation}
\hat{T}_{1}(r,\theta)=\left(\frac{C_{1}}{r^2}+C_{2}r\right)\cos{\theta},
\label{eq:16}
\end{equation}
where, upon using \eqref{eq:10} and \eqref{eq:15}, we have
\begin{align}
C_{1}(a,k,k_{r})&=\frac{a^2 \hat{T}_{0}(k-k_{r})}{2k+k_{r}}=\frac{a^2 \hat{T}_{0}(\Lambda-1)}{2\Lambda+1}, \label{eq:17}\\
C_{2}(a,k,k_{r})&=\frac{\hat{T}_{0}}{a}, \label{eq:18}
\end{align}
where $k_{r}=4 \epsilon \sigma \hat{T}_{0}^3 a$ is the effective radiative conductivity for the spherical inclusions, and $\Lambda=k/k_{r}$ plays the role of a conduction-to-radiation-ratio parameter. We note that this expression for $k_{r}$ can be obtained by linearising the Stefan-Boltzmann law for radiative heat flux and comparing it with a Fourier heat flux with an effective thermal conductivity $k_{r}$. Before we proceed, we note that \eqref{eq:16} with \eqref{eq:17}-\eqref{eq:18} is the unique solution for $\hat{T}_{1}$. The proof is non-standard, and we record it in Appendix A.


Having found $C_{1}$ in \eqref{eq:17}, we repeat the analysis from Section \ref{sec:2} to obtain that the coefficient for the far-field behaviour due to $n$ spheres is
\begin{equation}
n\delta C_{1}(a,k,k_{r})=\frac{k-k_{r}}{2k+k_{r}}\phi A^3 T'_{\infty},
\label{eq:24}
\end{equation}
which has the same functional form as \eqref{eq:241} with $k_{2}$ replaced with $k_{r}$ (we note that the factor of $\delta$ comes from the expansion \eqref{eq:12}). Thus, we find that the effective thermal conductivity is given by
\begin{equation}
k_{\mathit{eff}}=\frac{2k+k_{r}+2\phi(k_{r}-k)}{2k+k_{r}-\phi(k_{r}-k)}k=\frac{2\Lambda+1+2\phi(1-\Lambda)}{2\Lambda+1-\phi(1-\Lambda)}k.
\label{eq:27}
\end{equation}
We again note the immediate relation to \eqref{eq:3}, with $k_2$ replaced with $k_r$. This comes as no surprise given that we have linearised the radiative heat flux in \eqref{eq:9} assuming a small gas void in a uniform temperature gradient. Furthermore, note that if $\phi=0$, then $k_{\mathit{eff}}=k$, which is exactly as expected since this corresponds to the case when there is only the matrix material present. 

We also note that if $\Lambda=k/k_{r} \gg 1$ (the case where the radiative conductivity is much smaller than the solid one), then we obtain the asymptotic value for the effective thermal conductivity as
\begin{equation}
k_{\mathit{eff}}=\frac{2(1-\phi)}{2+\phi}k.
\label{eq:28}
\end{equation}
Similarly, if $\Lambda \ll 1$, then
\begin{equation}
k_{\mathit{eff}}=\frac{1+2\phi}{1-\phi}k.
\label{eq:29}
\end{equation}
These asymptotic scalings suggest that the effective thermal conductivity under the Maxwell model levels off to the constant value \eqref{eq:29} in the high-temperature regime.

\section{Effective medium theory with radiation}
\label{sec:4}

Having already calculated the relevant coefficients for the far-field behaviour due to the spherical inclusions with radiation in Section \ref{sec:3}, we straightforwardly generalise the EMT model presented in Section \ref{sec:2} to obtain
\begin{equation}\begin{aligned}
k_{\mathit{eff}} & = \frac{1}{4}\left\lbrace\vphantom{\sqrt{(3\phi (k_{r}-k)+(2k-k_{r}))^2}} 3\phi (k_{r}-k)+(2k-k_{r})\right.\\
& \qquad \left. +\sqrt{(3\phi (k_{r}-k)+(2k-k_{r}))^2+8k_{r}k}\right\rbrace \\
& = \frac{k}{4\Lambda}\left\lbrace\vphantom{\sqrt{(3\phi (k_{r}-k)+(2k-k_{r}))^2}} 3\phi (1-\Lambda)+(2\Lambda-1)\right.\\
& \qquad \left. +\sqrt{(3\phi (1-\Lambda)+(2\Lambda-1))^2+8\Lambda}\right\rbrace .
\label{eq:37}
\end{aligned}\end{equation}
Note again the expected similarities between \eqref{eq:37} and \eqref{eq:5}, with $k_r$ replacing $k_2$.

Asymptotic bounds can be obtained in either the large or small $k_r$ limits. If $k_{r} \ll 1$, for example (i.e., $\Lambda \gg 1$), then
\begin{equation}
k_{\mathit{eff}} \sim \left(1-\frac{3}{2}\phi\right)k.
\label{eq:38}
\end{equation}
This is valid for $\phi < 2/3$, which lies in our assumed region of dilute-porosity limit. If $k_{r} \gg 1$ (i.e, $\Lambda \ll 1$), then
\begin{equation}
k_{\mathit{eff}}=\frac{k}{1-3\phi},
\label{eq:39}
\end{equation}
which is valid for $\phi < 1/3$. We thus observe that the topology undergoes a percolation threshold. If we consider \eqref{eq:39} when radiation is large, then we see that, for porosity  $0 \leq \phi < 1/3$, the conductivity scales with $k$. Thus, for small enough porosity, the effective thermal conductivity under the EMT model levels off to the constant value given in \eqref{eq:39} as temperature increases, as was also true of the Maxwell model. 

\section{Differential effective medium theory with radiation}
\label{sec:5}

We use the analysis in Section \ref{sec:2} to generalise the differential effective medium theory and include radiation in the gas phase. This time, expanding our generalised Maxwell model result \eqref{eq:27} (which gives the solution in the dilute limit) for small $\phi$, we find
\begin{equation}
b(k,k_{r})=\frac{3(k_{r}-k)}{2k+k_{r}}.
\label{eq:45}
\end{equation}
After performing the integration in the analogue of \eqref{eq:444}, we obtain
\begin{equation}
\left(\frac{k_{\mathit{eff}}-k_{r}}{k-k_{r}}\right)^3\frac{k}{k_{\mathit{eff}}}=(1-\phi)^3,
\label{eq:46}
\end{equation}
or
\begin{equation}
\left(\frac{\Lambda k_{\mathit{eff}}/k-1}{\Lambda-1}\right)^3\frac{k}{k_{\mathit{eff}}}=(1-\phi)^3,
\label{eq:4612}
\end{equation}
which again is \eqref{eq:461} with $k_{2}$ replaced by $k_{r}$.

\begin{figure*}
\begin{center}
\includegraphics[width=0.38\linewidth]{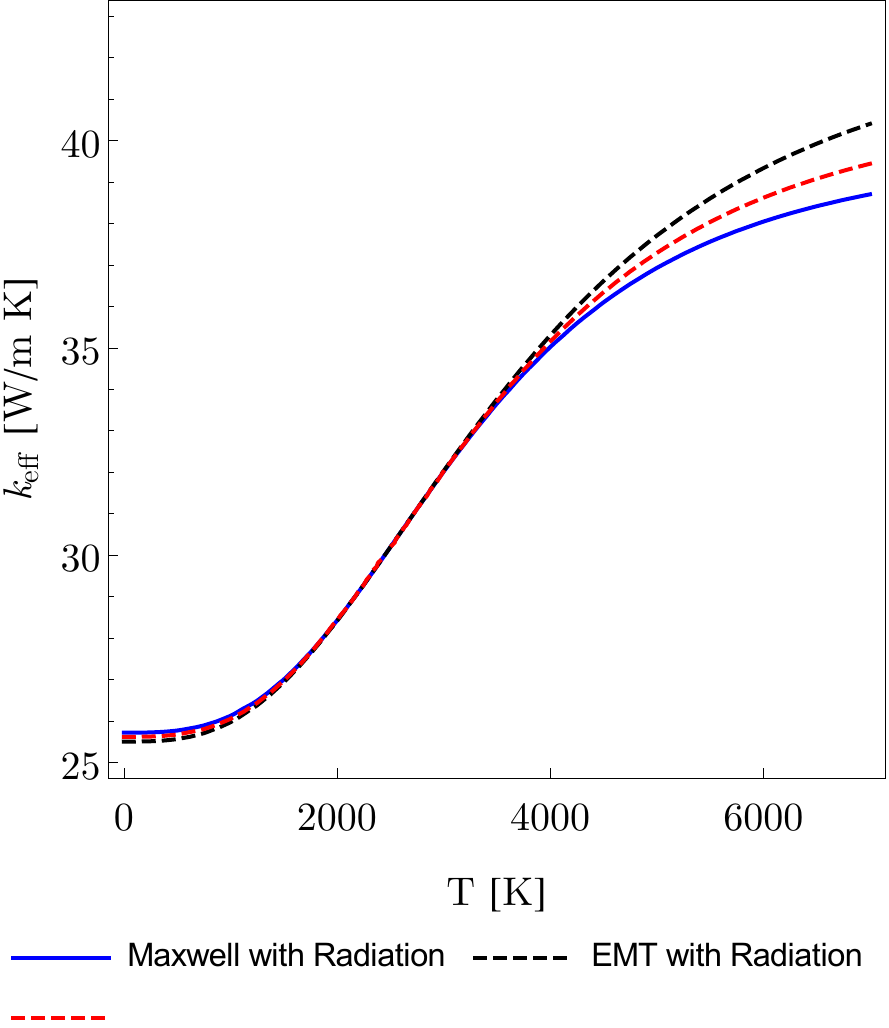}
\includegraphics[width=0.38\linewidth]{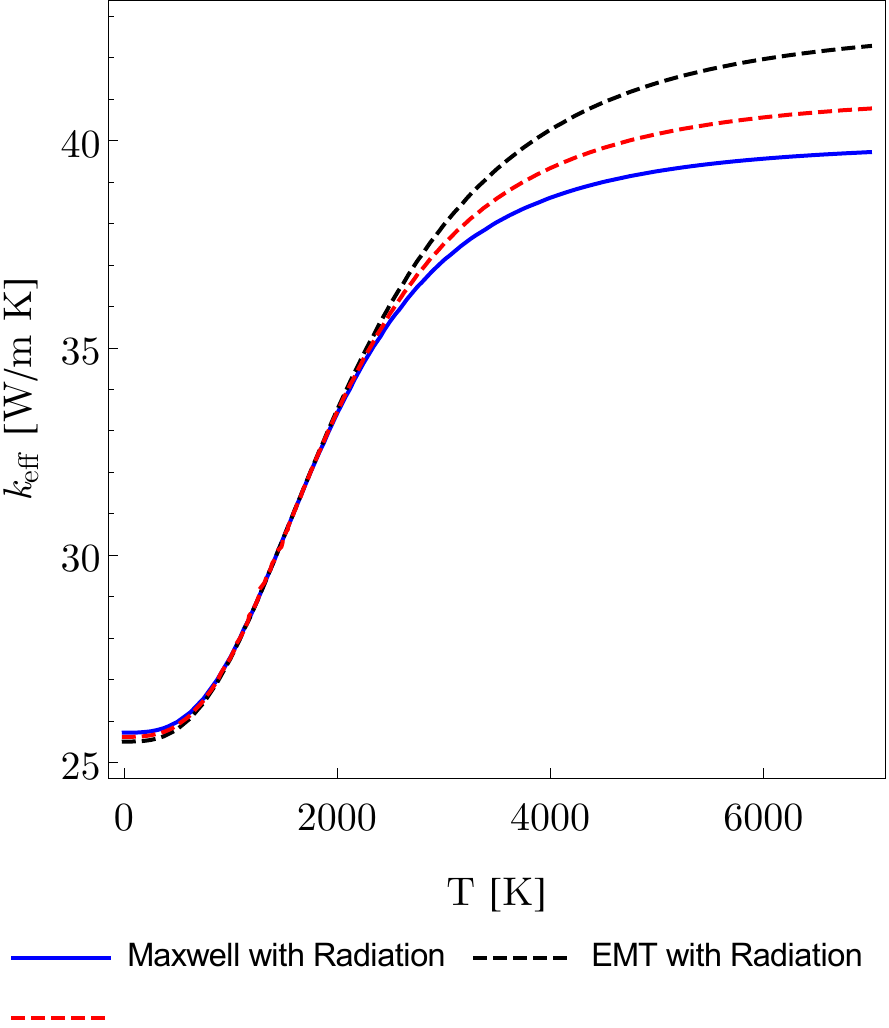}
\caption{Comparison between Maxwell, EMT, and DEMT models with gas voids of radius (a) $0.01$m, (b) $0.05$m.}
\label{fig:14}
\end{center}
\end{figure*}
We remark that this method relies on being able to incrementally change the medium, which is primarily applicable provided that there is a variety of gas void sizes. This is frequently the case in real-world applications and industrial processes.

As was done for previous models, asymptotic bounds may be obtained in the small or large $k_r$ limits. If $k_r \ll 1$ (i.e., $\Lambda \gg 1$), then we have
\begin{equation}
k_{\mathit{eff}}=(1-\phi)^{3/2}k,
\label{eq:47}
\end{equation}
while if $k_r \gg 1$ (i.e., $\Lambda \ll 1$), then we have
\begin{equation}
k_{\mathit{eff}}=\frac{k}{(1-\phi)^3}.
\label{eq:48}
\end{equation}
The DEMT theory can be seen as a perturbation of the dilute limit, and hence is valid for $\phi \ll 1$. Thus, for reasonable values of porosity, the effective thermal conductivity under the DEMT model will level off to a constant value \eqref{eq:48} in the high-temperature regime, as was true of the Maxwell and EMT models.

\section{Results and discussion}
\label{sec:6}

We compare numerically the three models for the effective thermal conductivity due to radiation within the gas voids. We choose parameter values corresponding to those used in \cite{Kosowska-Golachowska2014}, in order to calibrate our model to obtain physically meaningful results. We apply the models obtained in earlier sections to a solid matrix made of anthracite with gas pores inside. We use an anthracite conductivity of $30 \text{W}\text{m}^{-1}\text{K}^{-1}$, $\epsilon=1$, porosity of $\phi=0.1$, and average radii of gas inclusions either $0.01$m or $0.05$m. In Figure \ref{fig:14}, we compare results from the three models for the two different void radii by plotting the dependence of the effective thermal conductivity on temperature. The first thing to notice is that all three models give effective thermal conductivities which exhibit similar behaviours. The material with smaller gas pores results in greater agreement between the three effective thermal conductivities for a wider range of temperatures, as the radiation effect is less pronounced for smaller gas voids. 

All three models predict a saturation in the effective thermal conductivity at high temperatures. This is in contrast to many lumped parameter models, which often predict unbounded growth. The difference is partially due to the fact that there is a solid matrix with gas voids inside, therefore the solid is the rate-limiting factor. Now, comparing this observation with what is seen in the one-dimensional case, which corresponds to gas and solid blocks connected in series, we also find that $k_{\mathit{eff}}=1/(\phi/k_{r}+(1-\phi)/k)\to k/(1-\phi)$ when the temperature $T$ becomes very large. This is an interesting result, which says that if we have a porous solid material with gas bubbles inside (of small void fraction), the effective conductivity will eventually saturate with increasing temperature. This is in contrast to what one might expect if we have a composite material consisting of solid particles dispersed in a gaseous matrix, when the effective conductivity can be highly dependent on temperature (potentially unbounded).

For our given parameters, we also evaluate the conduction-to-radiation parameter to be $\Lambda = k/k_r \approx 0.5$ for radius of $0.01$m and $T=3000$ C, which shows that even for high-temperature regimes (such as those in industrial processes), solid conduction is comparable with radiative effects. This means that radiation does not become dominant. Of course, for voids of a larger radius, $\Lambda$ decreases. For lower temperatures of $T=1000$ C, we have that $\Lambda \approx 10$, while for higher temperatures of $T=5000$ C, we have that $\Lambda \approx 0.1$. Therefore, in these regimes we expect our asymptotic limits  \eqref{eq:28}, \eqref{eq:29}, \eqref{eq:38}, \eqref{eq:39}, \eqref{eq:47}, and \eqref{eq:48} may hold. Indeed, according to \eqref{eq:29}  we should have $k_{\mathit{eff}} \sim 40\text{W}\text{m}^{-1}\text{K}^{-1}$, and this is roughly the saturation value for the effective thermal conductivity (under the Maxwell model with radiation) we observe in Figure \ref{fig:14}.

\section{Conclusions}
\label{sec:7}

We have considered Maxwell's model, effective medium theory (EMT), and differential effective medium theory (DEMT) for the effective thermal conductivity of a porous material with small gas-filled voids in which radiation effects within the voids are taken into account. The corresponding expressions giving the relationship between the effective thermal conductivity and temperature, and parametrically depending on the thermal conductivity of the solid and the porosity of the material, are derived. The formulas for the effective thermal conductivity obtained under each model compare naturally with their original counterparts, if one were to use $k_{r}$ in place of the conductivity of the inclusions $k_{2}$, as is intuitively expected, justifying the use of the effective radiative conductivity in these models when considering the case of a solid matrix with gaseous inclusions. Furthermore, the standard formulas without radiation are obtained when the void fraction vanishes, providing another consistency check of our results. Our results justify the use of an effective radiative conductivity, obtained from linearising the Stefan-Boltzmann law. 

One interesting finding was that the predicted effective thermal conductivity saturates with increasing temperature; as the temperature increases, the rate of increase of the conductivity of the composite material slows down as the temperature increases, provided the porosity of the bulk material was not too large. Recall that in \cite{luikov1968thermal} there were a number of experimental results and scaling laws for the thermal conductivity of a range of porous materials, and at high temperatures thermal conductivities were observed to level off for some materials, rather than to increase without bound. (The effective thermal conductivities may have levelled off eventually for other materials, but the range of temperatures was limited in some cases, and was usually below what we considered in Figure \ref{fig:14}.) Such asymptotically bounded effective thermal conductivity is in contrast to what is observed in certain lumped parameter models, as those models often more crudely approximate the underlying physics. That this saturation occurs has important implications for a variety of industrial processes and may give insight into material design.


\section*{Acknowledgments}
This publication is based on work supported by the EPSRC Centre for Doctoral Training in Industrially Focused Mathematical Modelling (EP/L015803/1) in collaboration with Teknova, where it is part of the research project Electrical Conditions and their Process Interactions in High Temperature Metallurgical Reactors (ElMet), with financial support from The Research Council of Norway and the companies Alcoa, Elkem, and Eramet Norway. K. Kiradjiev thanks Teknova for financial support and the opportunity to work on-site during parts of this project. The authors thank C. M. Rooney and M. Sparta for helpful discussions.

\appendix

\section{Uniqueness of first-order correction, $\hat{T}_{1}$}

We prove that the first-order correction $\hat{T}_{1}$ to the temperature field in Maxwell's model with radiation is unique.

Suppose that there are two solutions for $\hat{T}_{1}$, say, $\hat{T}_{11}$ and $\hat{T}_{12}$. Let $\tilde{T}=\hat{T}_{11}-\hat{T}_{12}$. Then,
\begin{align}
\nabla^2\tilde{T}&=0 \qquad &&\text{ for } \qquad r > a, \label{eq:19}\\
\frac{k}{4 \epsilon \sigma \hat{T}_{0}^3}\dfrac{\partial \tilde{T}}{\partial r}&=\tilde{T}-\int_{\Sigma} \tilde{T} \mathrm{d}F \qquad &&\text{ on } \qquad r=a, \label{eq:20}\\
\tilde{T} &\to 0 \qquad &&\text{ as } \qquad r\to \infty, \label{eq:21}\\
\nabla \tilde{T} &= O(1/r^2) \qquad &&\text{ as } \qquad r\to \infty. \label{eq:22}
\end{align}
Using the Divergence theorem, consider the following:
\begin{equation}
\begin{aligned}
0&\leq \int_{\mathbb{R}\setminus V} k|\nabla\tilde{T}|^2\mathrm{d}V=\int_{\mathbb{R}\setminus V} k|\nabla\tilde{T}|^2\mathrm{d}V+\int_{\mathbb{R}\setminus V} k\tilde{T}\nabla^2 \tilde{T} \mathrm{d}V\\ &=\int_{\mathbb{R}\setminus V}\nabla\cdot (k\tilde{T}\nabla\tilde{T})\mathrm{d}V
=\int_{\Sigma} k\tilde{T}\nabla\tilde{T}\cdot(-\mathbf{n})\mathrm{d}S \\ &=-\int_{\Sigma} k\tilde{T}\dfrac{\partial \tilde{T}}{\partial r}\mathrm{d}S=4\epsilon\sigma \hat{T}_{0}^3\int_{\Sigma}\tilde{T}\left(-\tilde{T}+\int_{\Sigma} \tilde{T} \mathrm{d}F\right)\mathrm{d}S\\
&=4 \epsilon\sigma \hat{T}_{0}^3\left(-\int_{\Sigma}\tilde{T}^2 \mathrm{d}S+\frac{1}{4\pi a^2}\int_{\Sigma}\tilde{T} \mathrm{d}S\int_{\Sigma}\tilde{T} \mathrm{d}S\right)\leq 0,
\end{aligned}
\label{eq:125}
\end{equation}
where $V$ is the region $\{r \leq a \}$, $\mathbf{n}$ is the outwards-pointing unit normal to $\Sigma$, and the last inequality follows from the Cauchy-Schwarz inequality, \textit{viz}.,
\begin{equation}
\left(\int_{\Sigma}\tilde{T} \mathrm{d}S\right)^2\leq \int_{\Sigma} \mathrm{d}S\int_{\Sigma}\tilde{T}^2 \mathrm{d}S,
\label{eq:126}
\end{equation}
with $\int_{\Sigma} \mathrm{d}S=4\pi a^2$ being the measure (in this case, surface area) of the sphere. Again because of the Cauchy-Schwarz inequality, equality is possible only when $\tilde{T}$ and $1$ are linearly dependent, i.e., when $\tilde{T}$ is a constant. Using \eqref{eq:21}, we obtain $\tilde{T}=0$. Therefore, the solution for $\hat{T}_{1}$ is unique.

\end{document}